\documentclass[graybox]{svmult}

\usepackage{mathptmx}       
\usepackage{helvet}         
\usepackage{courier}        
\usepackage{type1cm}        
\usepackage{makeidx}         
\usepackage{graphicx}        
\usepackage{multicol}        
\usepackage[bottom]{footmisc}

\usepackage{booktabs}
\usepackage{multicol}

\makeindex             

\begin{document}

\title*{Enabling Dialogue Management with Dynamically Created Dialogue Actions}
\author{Juliana Miehle, Louisa Pragst, Wolfgang Minker, Stefan Ultes}
\institute{Juliana Miehle, Louisa Pragst, Wolfgang Minker, Stefan Ultes\at Institute of Communications Engineering, Ulm University, Germany}
%
%
\maketitle

\abstract*{In order to take up the challenge of realising user-adaptive system behaviour, we present an extension for the existing OwlSpeak Dialogue Manager which enables the handling of dynamically created dialogue actions. This leads to an increase in flexibility which can be used for adaptation tasks. After the implementation of the modifications and the integration of the Dialogue Manager into a full Spoken Dialogue System, an evaluation of the system has been carried out. The results indicate that the participants were able to conduct meaningful dialogues and that the system performs satisfactorily, showing that the implementation of the Dialogue Manager was successful.}

\abstract{In order to take up the challenge of realising user-adaptive system behaviour, we present an extension for the existing OwlSpeak Dialogue Manager which enables the handling of dynamically created dialogue actions. This leads to an increase in flexibility which can be used for adaptation tasks. After the implementation of the modifications and the integration of the Dialogue Manager into a full Spoken Dialogue System, an evaluation of the system has been carried out. The results indicate that the participants were able to conduct meaningful dialogues and that the system performs satisfactorily, showing that the implementation of the Dialogue Manager was successful.}

\section{Introduction}
\label{sec:Introduction}

One of the main challenges in Spoken Dialogue Systems is to realise effective dialogue strategies for coherent interactions and user-adaptive system behaviour. In general, the Dialogue Manager (DM) gets the user's input in form of a dialogue action which is a semantic representation of the user's utterance. Afterwards, it decides on the system's response based on the discourse context and outputs the semantic representation of the next system action. However, current DMs are often restricted to predefined dialogue actions leading to a loss in flexibility and robustness. Our aim is to increase the flexibility of the DM by the use of dynamically created dialogue actions in order to adapt the system's behaviour to the user, as proposed for example in~\cite{honold2014,miehleYoshinoPragstUltesNakamuraMinker2016,pragstMinkerUltes2017,ultesKrausSchmittMinker2015}. Thus, the conversation agent may appear more familiar and trustworthy and the dialogue may be more effective. 

In this work, we describe the implementation and evaluation of the extension of the already existing OwlSpeak DM~\cite{heinrothDenichSchmitt2010,ultesMinker2014} in order to handle dynamically created user and system actions, utilising general dialogue actions combined with ontology semantics to determine the system behaviour based on ~\cite{meditskosDasiopoulouPragstUltesVrochidisKompatsiarisWanner2016,pragstUltesKrausMinker2015}. The structure of the paper is as follows: In Section~\ref{sec:OwlSpeak}, the original OwlSpeak DM is introduced. Subsequently, we present our extension in order to handle dynamically created dialogue actions in Section~\ref{sec:OwlSpeakExtension} and an evaluation of the resulting system in Section \ref{sec:Evaluation}. In Section~\ref{sec:RelatedWork}, we discuss related work before concluding in Section~\ref{sec:Conclusion}.

\section{The OwlSpeak Dialogue Manager}
\label{sec:OwlSpeak}

\begin{figure}[ht]
	\begin{center}
		\includegraphics[width=0.9\textwidth]{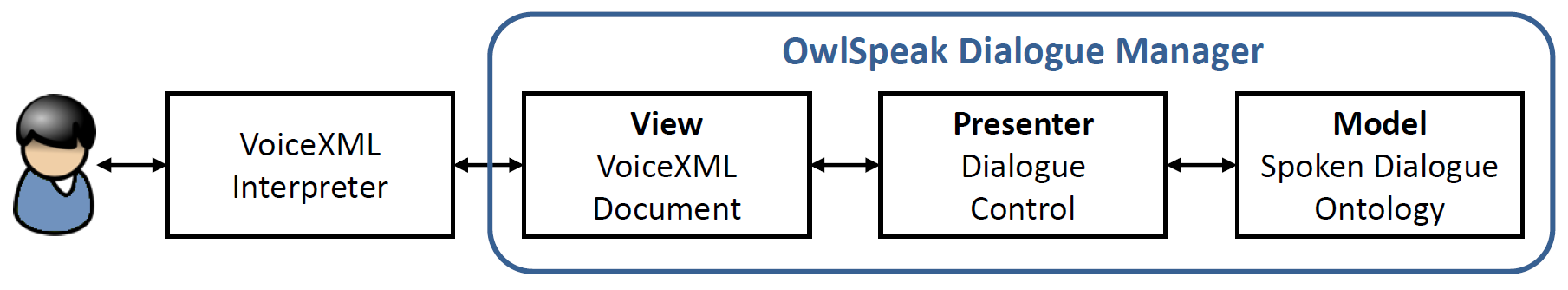}
		\caption{The architecture of OwlSpeak incorporating the Model-View-Presenter design pattern.}
		\label{fig:OwlSpeak_original}
	\end{center}
\end{figure}

In the following, the original OwlSpeak DM will be described based on the work by Ultes and Minker~\cite{ultesMinker2014}. OwlSpeak is an ontology-based DM which enables adaptive spoken dialogue within Intelligent Environments. The concept underlying Owl\-Speak incorporates the Model-View-Presenter design pattern~\cite{potel1996} whereby data management, dialogue logic and dialogue interface are separated, as can be seen in Figure~\ref{fig:OwlSpeak_original}.

\vspace{-0.5cm}
\noindent
\textbf{Spoken Dialogue Ontology (Model)} \hspace{0.1cm} The \textit{Model} is implemented as a Spoken Dialogue Ontology and consists of a static \textit{Speech} part which contains the concepts of the dialogue (e.g. pre-defined \textit{Grammar Moves} for the user and \textit{Utterance Moves} for the system) and a dynamic \textit{State} part which comprises concepts for the current state of the dialogue system (e.g. \textit{Agendas} representing one system action containing zero or one \textit{Utterance Moves} and one or more \textit{Grammar Moves} and the \textit{WorkSpace} storing all \textit{Agendas} that might be executed in following turns).

\vspace{0.2cm}
\noindent
\textbf{Dialogue Generation (Presenter)} \hspace{0.1cm} The \textit{Presenter} constitutes the dialogue control logic and thus the computational part of OwlSpeak and consists of a JAVA Servlet. It selects an \textit{Agenda} out of the \textit{WorkSpace}, creates a \textit{View} and then processes the user input that eventually is passed back by the \textit{View}.

\vspace{0.2cm}
\noindent
\textbf{Dialogue Interface (View)} \hspace{0.1cm} The \textit{View} is realised as a VoiceXML document which is created by the \textit{Presenter} and passed to the speech recogniser. There, the VoiceXML document is interpreted and output to the user. Then, the user input is passed back to the \textit{Presenter}.

\section{Handling Dynamically Created Dialogue Actions with OwlSpeak}
\label{sec:OwlSpeakExtension}

\begin{figure}[ht]
	\begin{center}
		\includegraphics[width=0.9\textwidth]{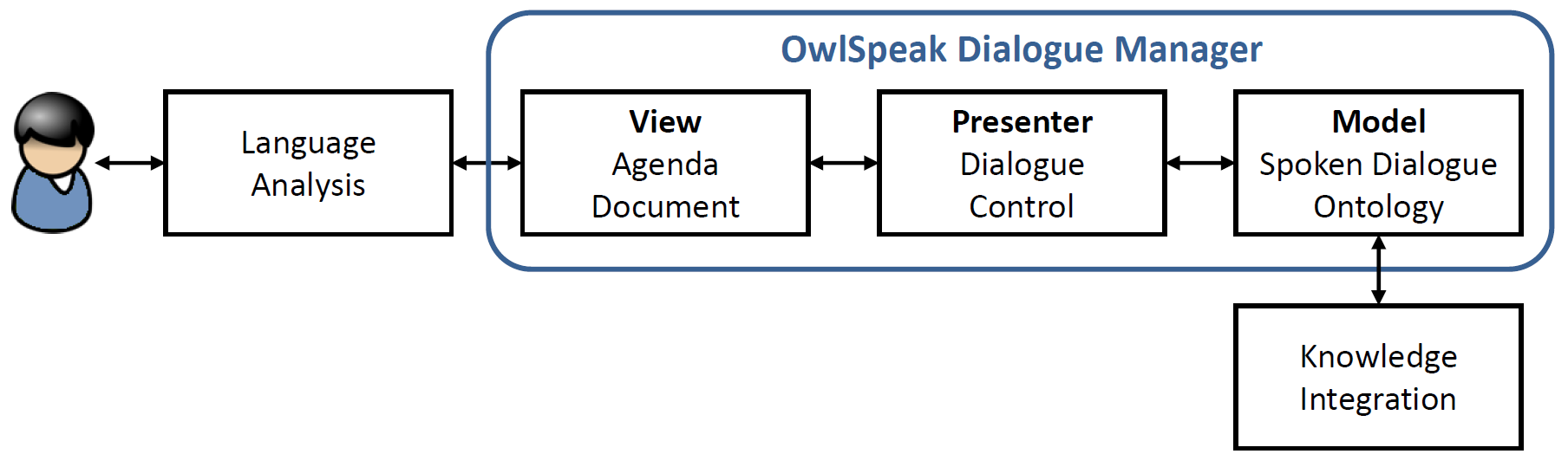}
		\caption{The modified architecture of OwlSpeak allowing to handle dynamically created actions.}
		\label{fig:OwlSpeak_extended}
	\end{center}
\end{figure}

As described by Meditskos et al.~\cite{meditskosDasiopoulouPragstUltesVrochidisKompatsiarisWanner2016}, modules for advanced techniques in the fields of language analysis as well as knowledge interpretation and reasoning need to be integrated in order to support dynamically created dialogue actions both by the user and the system. To facilitate the interaction with such modules, the \textit{Model} and the \textit{View} of OwlSpeak need to be adapted. The \textit{Presenter} including the dialogue control logic does not need to be modified as the general concept of selecting an \textit{Agenda} out of the \textit{WorkSpace} is still valid. The necessary modifications are depicted in Figure~\ref{fig:OwlSpeak_extended} and described in the following.

\vspace{-0.5cm}
\noindent
\textbf{Spoken Dialogue Ontology (Model)} \hspace{0.1cm} The \textit{Model} in form of a Spoken Dialogue Ontology needs to be modified in order to allow the communication with an external knowledge integration (KI) module which feeds the OwlSpeak DM with contextual information. This is implemented in such way that the external KI module sends relevant information that might be output to the user in the current situation. These information snippets are marked either as \textit{informable} or \textit{requestable}. Afterwards, OwlSpeak dynamically creates new \textit{Agendas} and stores them in the \textit{WorkSpace}. Each Agenda contains exactly one \textit{DialogueAction}, either \textit{request} or \textit{inform}, which is a new concept that has been added to the Spoken Dialogue Ontology. In contrast to a \textit{Move}, it does not contain any pre-defined \textit{Utterance}. However, the ontology semantics provided by the external KI module are also added to the \textit{Agenda}. Moreover, the \textit{Agendas} are assigned an \textit{age} indicating at which point of the dialogue they have been added to the \textit{WorkSpace}. This information might be used by the \textit{presenter} during the process of selecting the next \textit{Agenda} out of the current \textit{WorkSpace}. In addition to the dynamically created \textit{Agendas} based on the input of the external KI module, the \textit{WorkSpace} holds some \textit{Agendas} containing general \textit{DialogueActions} like \textit{greet}, \textit{acknowledge} and \textit{thank} which might be selected without asking the external KI module as they are self-contained and can be used in any dialogue domain.

\vspace{0.2cm}
\noindent
\textbf{Dialogue Interface (View)} \hspace{0.1cm} In order to allow a communication with an external language analysis module, a new interface has to be implemented which is based on a purely semantic representation of the user input and system output (rather than a sequence of fixed system utterances and corresponding user responds in form of grammars). Therefore, we introduced the \textit{Agenda Document} for the system output containing the selected \textit{Agenda} which should be performed, including its \textit{DialogueAction} and the corresponding ontology semantics in form of RDF Triples provided by the external KI module. The \textit{Agenda Document} is passed to the external language analysis module which extracts the semantic information and generates the corresponding system utterances. On the other hand, the language analysis module creates RDF Triples from the user input and passes them back to the \textit{Presenter} which extracts the \textit{DialogueAction} and decides on how to proceed.

\section{Evaluation}
\label{sec:Evaluation}

After implementing the presented extension in order to handle dynamically created dialogue actions with OwlSpeak and integrating the DM into the overall framework described in~\cite{kristina2017}, several evaluation sessions with human users were carried out. In total, 41 participants tested the system. The assessment was performed in accordance with the Guideline for Good Clinical Practice~\cite{gcp2016}. The procedure was as follows: First of all, the participants got a short introduction about the system as well as the functionalities. Moreover, they were informed about the evaluation process, the collected data and the internal use of data. Afterwards, each participant conducted a guided conversation with the system. At the end, a questionnaire had to be completed, containing statements about the overall system. Each statement had to be rated on a five-point Likert scale (1 = completely agree, 5 = completely disagree). Some results of the evaluation are shown in Table \ref{tab:eval}.

\begin{table}
\caption{Evaluation of the implemented system. The statements had to be rated on a five-point Likert scale (1 = completely agree, 5 = completely disagree).}
\label{tab:eval}
\centering
\begin{tabular}{l c c}
\toprule
 & Mean & Median \\
\midrule
The system returns sufficient information. & 2.88 & 3 \\
The system returns relevant information with the question. & 2.85 & 3 \\
The system returns reliable/trustworthy information. & 2.68 & 3 \\
The system returns meaningful responses. & 2.49 & 2 \\
\bottomrule
\end{tabular}
\end{table}

It can be seen that the participants were able to conduct meaningful dialogues. There is a tendency that our system returns sufficient information. The information seems to be quite relevant with the question as well as reliable and trustworthy. Moreover, the participants agree that the responses returned by the system are meaningful. We can conclude that, overall, the systems performs satisfactorily, even if there is still room for improvement. However, we think that the relevance, the meaningfulness as well as the reliability of the information does not only rely on the DM, but depends very much on the performance of the external knowledge integration module.

\section{Related Work}
\label{sec:RelatedWork}

There have been alternative approaches to separate the domain knowledge from the DM. The RavenClaw framework~\cite{bohusRudnicky2003} introduces a clear separation between task and discourse behaviour specification, allowing for a rapid development of DM components for goal-oriented domains. It consists of a Dialogue Task Specification layer which models the domain-specific dialogue logic, and a Dialogue Engine, which is domain-independent and executes the Dialogue Task Specification. Nothdurft~et~al.~\cite{nothdurftRichterMinker2014} present an architecture where a planner is used in order to provide explanations for the system's proposed course of action. The task-oriented dialogues are thereby modelled as a finite-state machine, while the planner outputs a decision tree. The DM compares the resulting actions and inserts pre-defined explanations when potential points of distrust are identified. The LS-SDS dialogue system~\cite{papangelis2017} is envisioned to support user requests over multiple, complex, rich, and open-domain data sources that will leverage the wealth of the available Linked Data. It is connected with an exploratory search system that supports the previously defined hard and soft restriction actions that allow the user to order facets, values, and objects. However, all of these systems are implemented in order to support slot-filling dialogues. In contrast, the approach presented in this work comprises both dialogues that consist of different slots that need to be filled and dialogues that are more chat-oriented and do not depend on any pre-defined slots. The scope and the topics of the dialogue depend on the external knowledge integration module, while the OwlSpeak DM can handle all kind of \textit{informable} or \textit{requestable} information.

\section{Conclusion}
\label{sec:Conclusion}

We presented the implementation of an extension for the existing OwlSpeak DM to enable the handling of dynamically created dialogue actions. This leads to an increase in flexibility which can be used for adaptation tasks. In order to support dynamically created dialogue actions both by the user and the system, modules for advanced techniques in the fields of language analysis as well as knowledge interpretation and reasoning need to be integrated. To facilitate the interaction with these modules, the \textit{model} and the \textit{view} of OwlSpeak needed to be adapted. After the implementation of the modifications and the integration of the DM into the overall framework, an evaluation of the system has been presented. The results indicate that the participants were able to conduct meaningful dialogues and that the system performs satisfactorily, showing that the implementation of the DM was successful.

\begin{acknowledgement}
This work is part of a project that has received funding from the \textit{European Union's Horizon 2020 research and innovation programme} under grant agreement No 645012. We thank our colleagues from the University of T\"ubingen, the German Red Cross in T\"ubingen and semFYC in Barcelona for organizing and carrying out the evaluation.
\end{acknowledgement}

\bibliographystyle{spmpsci}
\bibliography{references}
\end{document}